\documentclass[floats,aps,twocolumn]{revtex4}
\usepackage{graphicx}
\usepackage{color}
\usepackage{amsmath}
\begin{document}
\newcommand{\rr}{{\bf r}}
\newcommand{\kk}{{\bf k}}
\newcommand{\pp}{{\bf p}}
\newcommand{\qq}{{\bf q}}
\newcommand{\Q}{{\bf Q}}
\newcommand{\BSCCO}{{Bi$_2$Sr$_2$CaCu$_2$O$_8$ }}
\newcommand{\YBCO}{{YBa$_2$Cu$_3$O$_{7-\delta}$ }}
\def\k{{\bf k}}
\def\r{{\bf r}}
\def\R{{\bf R}}
\def\p{{\bf p}}
\def\D{{\cal D}}
\def\q{{\bf q}}
\def\G{{\hat{G}}}
\def\V{{\hat{V}}}
\def\T{{\hat{T}}}

\title{Power spectrum of many impurities in a $d$-wave superconductor}
\author{Lingyin Zhu$^{1}$, W. A. Atkinson$^2$, and
P. J. Hirschfeld$^{1}$\\~}

\affiliation{$^1$Department of Physics, University of Florida,
Gainesville FL 32611\\$^2$Department of Physics, Southern Illinois
University, Carbondale IL 62901-4401 }
\date{\today}

\begin{abstract}
Recently the structure of the measured local density of states
power spectrum of a small area of the \BSCCO (BSCCO) surface has
been interpreted in terms of peaks at an ``octet" of scattering
wave vectors determined assuming weak, noninterfering scattering
centers. Using analytical arguments and numerical solutions of the
 Bogoliubov-de Gennes equations, we discuss how the
interference between many impurities in a  $d$-wave superconductor
alters this scenario.  We propose that the peaks observed in the
power spectrum are not the features identified in the simpler
analyses, but rather  ``background" structures which disperse
along with the octet vectors.  We further consider how our results
constrain the form of the actual disorder potential found in this
material.

\end{abstract}

\maketitle

In the past few years, high-resolution scanning tunneling
microscopy (STM) experiments on the cuprate superconductor BSCCO
\cite{yazdani,davisnative,davisZn,cren,davisinhom1,davisinhom2,Kapitulnik1,Kapitulnik2,Kapitulnik3,Hoffman1,McElroy}
have obtained local information on electronic structure for the
first time.  The first great success of this technique was the
observation of resonant defect states at low temperatures in the
superconducting state\cite{yazdani,davisnative,davisZn},
confirming early proposals that such states should be reflected in
the local density of states (LDOS) of $d$-wave
superconductors\cite{Byersetal,Balatsky}. Subsequent experiments
revealed the existence of nanoscale
inhomogeneities\cite{cren,davisinhom1,davisinhom2,Kapitulnik1}
which are currently the subject of debate, being attributed either
to interaction-driven effects such as
stripe-formation\cite{Kapitulnik2,Kapitulnik3} or to Friedel
oscillations of weakly-interacting
quasiparticles\cite{davisinhom1,davisinhom2}.  At the heart of
this debate lies the important question of whether, and if so in
what ranges of doping, conventional BCS-like theories describe the
ground state and low energy excitations of BSCCO.  Recently it has
been pointed out that, in inhomogeneous systems, the Fourier
transform of the LDOS (FTDOS) contains information not only about
the disorder potential, but about the kinematics of the associated
pure system. That this must be so at some level is clear from the
original one-impurity problem solved by Friedel\cite{friedel}: a
charge inserted in an electron gas gives rise to LDOS oscillations
which vary at large distances as $\sim \cos 2k_F r/r^3$, so the
wavelength of the LDOS ``ripples" caused by a single impurity
gives the Fermi wave vector directly.  A somewhat more
sophisticated version of this
argument,\cite{Hoffman1,Tingfourier,WangLee,Hedegard,Tingfourierfewimp}
still assuming scattering from a single or few impurities and
noninteracting quasiparticles, suggested that the peaks in the
FTDOS are due to scattering of quasiparticles by a weak disorder
potential.  In this case favored momentum transfers correspond to
vectors ${\bf q}$ connecting two tips of quasiparticle constant
energy contours which maximize the joint density of states (JDOS).
This interpretation has been applied to recent
experiments\cite{McElroy} which claim to map out a Fermi surface
in agreement with angle-resolved photoemission (ARPES).  In
contrast to this, Howald et al.\cite{Kapitulnik2,Kapitulnik3}
identify nondispersing features in the FTDOS from their
experiments, which they suggest are indicative of static-stripe
formation along the Cu-O bond directions.

In this Letter, we report on numerical studies of models of
disordered superconducting BSCCO.  Previous analyses of a single
weak impurity\cite{Tingfourier,WangLee} agreed qualitatively with
the simple JDOS analysis of Ref.~\cite{McElroy} and succesfully
predicted the dispersion of the peak positions in the FTDOS. It
has been since shown\cite{capriotti} that the sharp 1-impurity
peaks survive in the many-impurity case if the potential is
sufficiently weak. However, the ability to correctly predict the
peak positions (which depend only on the quasiparticle band
structure, and not the disorder potential) does not imply that the
low-energy excitations are understood.  A full microscopic
understanding of the superconducting state requires an accurate
description of the  disorder potential in the BSCCO CuO$_2$
planes.  We will show that there are both qualitative and
quantitative problems describing experiments which stem from the
inadequacy of the weak-impurity model, and that the FTDOS spectrum
of a realistic model comprising a dilute concentration of unitary
scatterers, together with  a smooth disorder potential, is
required to fit experiment and has features not present in the
simple JDOS analysis. Understanding
 these discrepancies will be crucial to extracting reliable information
about  the clean system, as well as the nature of the disorder
potential.

\begin{figure}
\begin{center}
\includegraphics[width=\columnwidth]{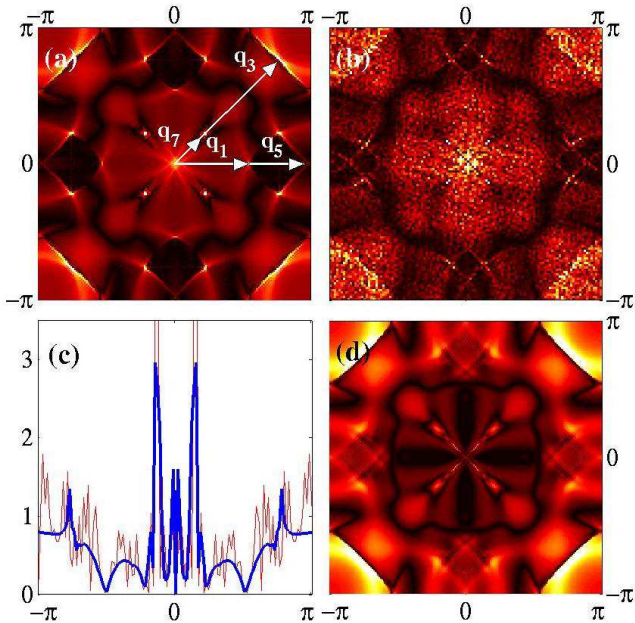}
\end{center}
\caption{FTDOS at $\omega=$14 meV for weak potential scatters
($V_0=0.67t_1$): (a) for one weak impurity, with a few important
scattering wavevectors indicated; (b) for 0.15\% weak scatterers.
Cuts through the data of (a)(thick line) and (b)(thin line) along
the (110) direction and scaled by $1/\sqrt{N_I}$ are plotted vs.
$q_x$ in (c), while (d) shows the weak scattering response
function Im $\Lambda_3(\q,\omega)$. Peaks at $\q = 0$ are removed
for clarity. In all figures, the $x$ and $y$ axes are aligned with
the Cu-O bonds. } \label{ftdos}
\end{figure}

{\em Many Impurities.} In this section, we discuss the differences
between the single and many-impurity FTDOS for weak pointlike
scatterers, and compare these spectra with experiment.  We first
imagine a random distribution of $N_I$ pointlike impurities
located at sites $\R_i$ with $i=1\ldots N_I$.  The LDOS can be
formally decomposed $\rho(\r,\omega) = \rho_0(\omega) + \delta
\rho(\r,\omega)$ where $\rho_0$ is the DOS of the homogeneous
superconductor, and $\delta \rho$ is the local shift due to
disorder given exactly by
    \begin{eqnarray}
    \delta \rho(\r,\omega) &=& -\frac{1}{\pi} \mbox{Im}
    \sum_{i,j=1}^{N_I} \left [\hat G^0(\r-\R_i)\hat T_{ij}\hat G^0(\R_j-\r)
    \right ]_{11}
    \label{drho}
    \end{eqnarray}
where the $\omega$-dependence is suppressed for clarity, $\hat
T(\omega)$ is the $2N_I\times 2N_I$ many-impurity T-matrix (the factor
of two arises from spin), $\hat G^0(\r,\omega)$ is the bare electron
Green's function $\hat G^0(\r,\omega) = \sum_\k \hat G^0(\k,\omega)
\exp(i\k\cdot \r)$, $\hat ~$ refers to matrices in the Nambu-spinor
formalism and $[\ldots]_{ij}$ are spinor indices.
The T-matrix is expressed in terms of the 1-impurity T-matrix
$\hat t_i = [1-\hat V_i \hat G^0(\r=0)]^{-1}\hat V_i$ by
    \begin{eqnarray}
    \hat T_{ij} &=& \hat t_i \delta_{i,j} +
    \sum_{m=1}^{N_I} \hat t_i [1-\delta_{m,i}]\hat G^0(\R_i-\R_m) \hat
    T_{mj},\label{manyT}
    \end{eqnarray}
where the impurity potential at $\R_i$ is $\hat V_i = V_0\hat
\tau_3$, and $\hat \tau_i$ are the Pauli matrices.  For pure
potential scatterers, the T-matrix can be decomposed into two
Nambu components $\hat t = t_0e^{i\phi_0}\hat \tau_0 +
t_3e^{i\phi_3}\hat \tau_3$ with $t_0$ and $t_3$ real, and $\phi_0$
and $\phi_3$ the scattering phase shifts.

Throughout this work, we adopt the ARPES-derived tight-binding
model of Norman\cite{Norman} for the band structure, and assume
nearest-neighbor $d$-wave pairing with order-parameter $\Delta_\k
= \Delta_0(\cos k_x - \cos k_y)$ and $\Delta_0 = 0.16t_1 = 24$ meV
where $t_1 = 150$ meV is the nearest-neighbor hopping matrix
element.  We take, as representative, the weak scattering
potential $V_0 = 0.67 t_1$.  We ignore the effects of tunneling
matrix elements\cite{Tingfilter,Balatskyfilter}, and assume that
the STM probe measures the LDOS $\rho(\r,\omega)$ directly (note
that $\r$ need not correspond to a site $\R$ of the crystal
lattice).  The Fourier transform is then $\rho(\q,\omega) =
\sum_{\r\in L\times L} e^{-i\q\cdot \r} \rho(\r, \omega)$, where
$L\times L$ is a square set of $L^2$ positions at which
measurements are made, and $\q = 2\pi(m,n)/L$ are vectors in the
associated reciprocal lattice.

The FTDOS is shown at $\omega = 14$ meV in Fig.~\ref{ftdos} for (a) a
single impurity\cite{wangleecaveat} (b) a collection of $N_I=21$
(0.15\%) weak scatterers on a $120\times 120$ lattice.  Using the
notation of Ref.~\cite{McElroy}, a few important peak positions which
are predicted by the JDOS analysis are also shown in
Fig.~\ref{ftdos}(a).  To understand the relationship between
Figs.~\ref{ftdos}(a) and (b), we consider Eq.~(\ref{drho}) to leading
order in $\hat t_i$:
\begin{eqnarray}
\delta\rho (\q,\omega) &\simeq&
-\frac{1}{\pi}\sum_{\alpha=0,3} t_\alpha(\q) {\rm Im} [ e^{i\phi_\alpha}
 \Lambda_\alpha (\q,\omega)]
\label{separate}
\end{eqnarray}
where $t_\alpha(\q) = t_\alpha\sum_i e^{-i\q\cdot \R_i}$ and
$\Lambda_\alpha(\q,\omega)=
\sum_\k \left[
\G^0(\k,\omega)\hat \tau_\alpha \G^0(\k+\q,\omega)\right]_{11}.\nonumber
$
Eq.~(\ref{separate}) affords a clear separation between degrees of
freedom associated with the disorder potential and those of the
pure system, which is not possible if scattering processes of
higher order in $\hat t_{i}$ are important.  Since $t_\alpha(\q)$
consists of a sum of $N_I$ random phases, it becomes a random
function of $\q$ in the first Brillouin zone as
$N_I\rightarrow\infty$. By contrast, $\Lambda_\alpha(\q,\omega)$
is the response function of the clean system, is independent of
the disorder potential, and determines the peak locations and
widths in the FTDOS.  As is evident in Fig.~\ref{ftdos}(c), JDOS
peaks are {\it not broadened} or shifted by disorder; note in
particular that the thin line representing the many-impurity case
shows a sharp $\q_7$ peak which is visible only as one or two
bright pixels in Fig.~\ref{ftdos}(b).

  In the limit of weak potentials, Eq. \eqref{separate} reduces to the
result of Capriotti et al\cite{capriotti},
\begin{equation}
\delta \rho(\q,\omega) \simeq -V(\q) \mbox{Im
}\Lambda_3(\q,\omega)/\pi, \label{cap}
\end{equation}
which is also valid for finite range $V(\r)$. The effect of an
extended potential can be understood with an example: if the
single-impurity potential is $V_0(\r)=v_0 \exp(-r^2/2r_0^2)$, then
$V(\q) = 2\pi r_0^2 v_0 \exp(-q^2r_0^2/2) \sum_{i} \exp(i\q\cdot
\R_i)$. Thus, for sufficiently weak scatterers, the $q$-space
structure of $\delta \rho(\q,\omega)$ is determined primarily by
the band structure, but has an envelope which suppresses the FTDOS
near the Brillouin zone edges, and is noisy because of randomness
in the impurity positions. We remark that since
$\Lambda_3(\q,\omega)$ can be calculated from knowledge of the
electronic structure of the pure system, Eq.~(\ref{cap}) is, in
principle, a useful tool for determining the scattering potential
directly from experimental measurements of the LDOS.  However, we
caution the reader that Eq.~(\ref{cap}) applies only to
unphysically weak potentials. Figure~\ref{ftdos}(d) shows that
even for the weak potential $V_0=0.67t_1$, not all peaks in the
FTDOS (notably the central peak around $q=0$) are reproduced by
Eq.~(\ref{cap}).

While Eqs.~(\ref{separate}) and (\ref{cap}) appear to suggest that
the dispersing 1-impurity peaks are also present for many
impurities, we can show from Eq.~(\ref{separate}) that in the
disorder average $\langle |\delta \rho(\q,\omega)|^2 \rangle \sim
N_I$ and $\langle |\delta \rho(\q,\omega)|^4\rangle -\langle
|\delta \rho(\q,\omega)|^2 \rangle^2 \sim N_I(N_I-1)$, implying
that the noise is actually as large as the signal.  As a
consequence, the weak $\q_1$ peaks present in Fig.~\ref{ftdos}(a)
are lost in the noise in Fig.~\ref{ftdos}(b), and  the broader
background features are relatively enhanced.  Thus, it appears
that the peaks predicted by the JDOS analysis may not be robust
when many impurities are present.  This is particularly true when
the effects of energy resolution $\Delta \omega$ are considered.
In Ref.~\cite{McElroy}, $\Delta \omega \approx 2$
meV\cite{Davisprivatecommun}, and we use a complex energy
$\omega+i\gamma$ in our calculations which for $\gamma = 0.015t_1
= 2.25$ meV yields a comparable resolution.  We find that even for
one impurity, FTDOS peaks are extremely sensitive to $\gamma$. All
peaks are suppressed, {\em but not broadened}, as $\gamma$
increases (this conclusion differs from Ref.~\cite{WangLee}).

It is therefore difficult to understand, based on the weak
impurity analysis, why experiments measure {\em broad} and
well-defined $\q_1$ and $\q_7$ peaks of roughly equal weight.
Furthermore, 1-impurity calculations such as in
Fig.~\ref{ftdos}(c) find in addition to JDOS peaks a highly
structured dispersive ``background" structure, whereas in current
interpretations of experiments all features are ascribed to JDOS
peaks.   Our calculations suggest that, because of noise, it may
be easy to confuse dispersing background features and JDOS peaks
in experiments. Very recently, it was pointed out in Ref.
\cite{Franz} that the scattering amplitude for processes involving
$\Lambda_0(\q)$ is not sharply peaked at the ``octet" $\q_\alpha$,
but our analysis shows further that the predominant structures in
the presence of noise are generally considerably shifted from the
$\q_\alpha$ positions, and may indeed be associated with off-shell
processes.

{\em Realistic Disorder Models.}  The weak-scattering pointlike
disorder model is convenient for its simplicity, but does not
predict FTDOS peak widths consistent with experiment.  In
addition, the relative weights of the features ascribed to the
$\q_\alpha$  (e.g., the nearly equal weight of $\q_7$ and $\q_1$)
are not correctly predicted, particularly at low energies.
In this section, we explore whether the measured FTDOS can be
explained by a simple ``realistic'' disorder model.  Two types of
disorder, unitary defects and intrinsic nanoscale inhomogeneties
are known to be present in nominally clean BSCCO.  In the
experiments of Ref.~\cite{McElroy}, a concentration of $\approx
0.2\%$ unitary, pointlike native defects were observed, while in
Refs.~\cite{cren,davisinhom1,davisinhom2,Kapitulnik1,Kapitulnik2,Kapitulnik3}
nanoscale inhomogeneities were observed with a typical size of
$\approx 2$ nm.  In this work, we model the native defects as
pointlike scatterers with an on-site potential $V_0 = 30t_1$ which
produces a local resonance centered at 0 meV (the concentration
and strength of the unitary defects is thus fixed by experiment
within the potential scattering model).  The source of the
nanoscale inhomogeneity is unknown, and we take the simplest
ansatz, that it arises from a smooth random potential, probably
originating from charge inhomogeneities in the BiO layers.  For
definiteness, we model the smooth potential as $V(\r) = \sum_{i}
V(i) \exp(-\tilde r_i/\lambda) /\tilde r_i$ and $\tilde r_i =
[(\r-\R_i)^2 + d_z^2]^{1/2}$, where $\R_i+ {\bf \hat z}d_z$ are
the defect locations, $V(i)$ are the defect potentials and
$\lambda$ is a screening length.
We take a bimodal distribution $V(i) = \pm V$ so that the
smooth potential represents spatial fluctuations of the local
potential about a mean which is determined by the doping level.
Equation (\ref{manyT}) applies to pointlike impurities only,
so we use an implementation of the recursion method\cite{haydock}
to solve directly for the local Green's function $\hat
G(\r,\r,\omega+i\gamma)$, with $\gamma = 2.25$ meV, of the
inhomogeneous system.  Our best-fit model consists of 0.2\%
unitary defects, and 8\% smooth scattering centers with $V=2t_1$,
$d=2a$ and $\lambda=a$.

\begin{figure}[tb]
\begin{center}
\includegraphics[width=\columnwidth]{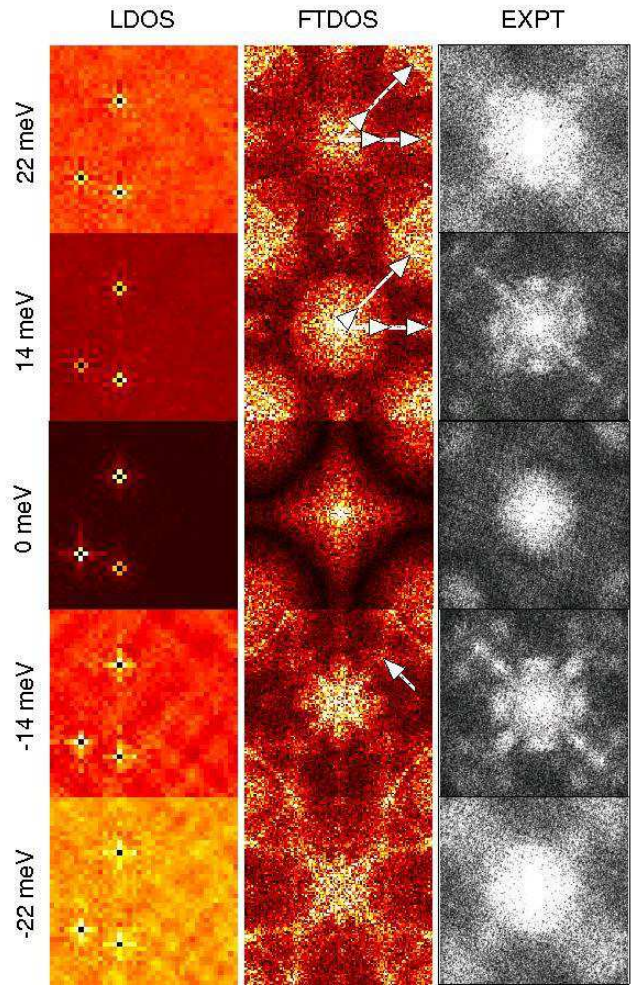}
\caption{ Comparison of theory and experiment.  Calculated LDOS
and FTDOS are compared with experimental FTDOS (EXPT) at five
energies. LDOS are shown for a $40\times 40$ subsection of the
$120\times 120$ lattice.  FTDOS are shown in the first Brillouin
zone, $-\pi/a \le q_x,q_y \le \pi/a$.  Theoretical calculations
are for a mix of unitary pointlike and smooth potential
scatterers.  Scattering $q$-vectors are shown at 22 meV and 14 meV
(negative energy vectors are the same).  A broad background
feature in the FTDOS is identified with an arrow at -14 meV.}
\label{fig3}
\end{center}
\end{figure}

We compare this mixed-impurity model with the experimental data of
McElroy et al in Fig.~\ref{fig3}.  In both experiment and
calculations, the local resonances are plainly evident in the LDOS
for $|\omega| \lesssim 15$ meV and, in our calculations, make the
dominant contribution to the FTDOS As with weak scatterers, the
narrow JDOS peaks are swamped by noise, and suppressed by the
finite energy resolution.  However, unlike the weak scattering
case, the $\q_1$ and $\q_5$ structures at 14 meV are relatively
robust and remnants of the peaks can be seen. Excess weight
relative to experiment in the Brillouin zone corners is probably
due to finite potential range and sub-unit cell information
obtained in experiment, both of which lead to a modulation of
$\delta \rho(\q)$ which decays at large $q$\cite{unpublished}.

For larger energies, the LDOS impurity resonances disappear in
experiment but, because the electronic wavefunctions are excluded
from the defect site, they remain clearly visible in our
calculations.  The discrepancy may arise because the STM tip
height is adjusted to maintain a constant current at a fixed bias
voltage (typically -150 meV in Ref.~\cite{McElroy}), so that what
is actually plotted is the LDOS relative to the average LDOS at
that site.  We will discuss the effect of tip-height adjustments
elsewhere\cite{unpublished}, and here we simply note that despite
discrepancies, the Fourier transforms of the measured and
predicted patterns are qualitatively similar.  At energies
$|\omega| \gtrsim 15$ meV, the smooth random potential makes a
significant contribution to the FTDOS.

The agreement between the calculated and experimental FTDOS is
qualitatively good at low energies, although our calculations show
more asymmetry between positive and negative energies, with the
negative energy calculations fitting better, than in experiments.
This is the result of the large, probably artificial, asymmetry of
the model band, in which there is a van Hove singularity at
$\omega \approx -50$ meV that is not observed in tunneling
experiments. As before, remnants of JDOS peaks (for example, the
$\q_1$ peak at -14 meV or the $\q_5$ peak at 14 meV) are evident
in the calculated FTDOS, but most of the structure comes from a
set of broader dispersing features, which has been previously
ignored. At -14 meV and -22 meV, we see that there are clear octet
structures about the central peak, but that the dispersing
``$\q_7$'' features [the arms along the ($\pm\pi,\pm\pi$)
directions extending from the central peak] extend well beyond
$\q_7$.  In general, the background structures disperse
qualitatively (growing or shrinking with $|\omega|$) as one
expects from the JDOS analysis.

Since the $\q_7$ peak comes from intranodal scattering, it is a
direct measure of the $\k$-dependence of the superconducting gap
and scales with $\sim 1/{v_\Delta}$, where ${v_\Delta}$ is the gap
velocity at the nodes. The measured gap is not consistent with
nearest-neighbor $d$-wave pairing---for example the experimental
value at $\omega=-14$ meV $\q_7^{exp}\approx 0.6
\pi/a$\cite{McElroy} is nearly twice the value in our
nearest-neighbor $d$-wave model---and the
authors of Ref.~\cite{McElroy} have been forced to introduce a
significant subleading $\cos 6 \theta$ harmonic in their fit of the
angular dependence of the gap on the Fermi surface.  The additional
harmonic is associated with pairing beyond near-neighbors, in contrast
to ARPES results which suggest the gap is pure near-neighbor $d$-wave
at optimal doping.
Recognizing that the observed feature at roughly twice the true
$\q_7$ is in fact the background feature (similar to that seen as
a ``hump" in Fig. 1(c)) found in our calculations may enable one
to bring the two experiments in closer agreement.

At $\omega = \pm 22$ meV, the calculated (110) structures, which
correspond to forward (intranodal) scattering, are stronger than
the $(100)$ structures.
 Preliminary numerical calculations suggest that
scattering from order parameter fluctuations may contribute to
$\q_1$-type peaks, leading to the interesting question of how the
interplay of disorder and order parameter fluctuations manifests
itself in the FTDOS. It is a general feature of our calculations
at higher energies that the $\q_1$ peaks are weaker than  observed
in experiments, and  a tendency to  stripe formation along the
Cu-O bonds, as proposed in
\cite{Kapitulnik1,Kapitulnik2,Kapitulnik3}, could possibly enhance
the weight near the $\q_1$ peaks.  However, stripe formation is
difficult to reconcile with  the dispersal of the $\q_1$
peak\cite{Hoffman1,McElroy}.

 {\it Conclusions.}  Existing analyses of recent STM measurements of
the FTDOS have established the fundamental point that the
structures in the FTDOS are dispersive and therefore likely
represent interference of disordered quasiparticle wavefunctions.
We have argued, using a combination of analytical and numerical
approaches, that previous weak-scattering analyses are inadequate
to explain the details of the FTDOS, however.  In previous works
it was expected that sharp ``octet" peaks would broaden into the
observed structures due to impurity scattering; here we have shown
that these peaks are lost in the noise created by the interference
of many impurities, rather than broadened. Our work suggests that
the dispersing features observed may not correspond directly to
the predicted sharp peaks at all, but rather to dispersive
background structures which have hitherto been ignored. We show
that the simplest ``realistic'' disorder model of unitary
impurities and a smooth random potential explains many features of
the FTDOS at lower energies, but at higher energies the comparison
is worse, probably reflecting our lack of knowledge of the true
smooth component of the disorder, including the local order
parameter fluctuations neglected here.   Further refining of these
comparisons will be extremely important in understanding the
origin of the nanoscale inhomogeneities observed in STM
experiments.

{\it Acknowledgements.} The authors would like to thank D.J.
Scalapino, R. Sedgewick, J.C. Davis and K. McElroy for several pivotal
discussions. WAA would like to acknowledge Research Corporation Grant
CC5543.


\begin{thebibliography}{scalapino}
\bibitem{wangleecaveat} The 1-impurity FTDOS is somewhat different
than in Ref.~\protect\cite{WangLee} because of the larger value of
$\Delta_0$ used here.
\end{thebibliography}
\end{document}